\documentclass[twoside,11pt]{jmlr}

\begin{document}

\title{Optimization Assisted MCMC}

\author{\Name{Ricky Fok} \Email{ricky.fok3@gmail.com} \\
       \addr Department of Electrical Engineering \& Computer Science and Engineering\\
       York University\\
       Toronto, Ontario, Canada, M3J 1P3
       \AND
       \Name{Aijun An} \Email{aan@cse.yorku.ca} \\
       \addr Department of Electrical Engineering \& Computer Science and Engineering\\
       York University\\
       Toronto, Ontario, Canada, M3J 1P3
       \AND
       \Name{Xiaogang Wang} \Email{stevenw@mathstat.yorku.ca} \\
       \addr Department of Mathematics and Statistics\\
       York University\\
       Toronto, Ontario, Canada, M3J 1P3}

\editor{???}

\maketitle

\begin{abstract}
Markov Chain Monte Carlo (MCMC) sampling methods are widely used but often encounter either slow convergence or biased sampling when applied to multimodal high dimensional distributions. In this paper, we present a general framework of improving classical MCMC samplers by employing a global optimization method. The global optimization method first reduces a high dimensional search to an one dimensional geodesic to find a
starting point close to a local mode. The search is accelerated and completed by using
a local search method such as BFGS. We modify the target distribution by extracting a
local Gaussian distribution aound the found mode. The process is repeated to find all
the modes during sampling on the fly. We integrate the optimization algorithm into the
Wormhole Hamiltonian Monte Carlo (WHMC) method. Experimental results show that,
when applied to high dimensional, multimodal Gaussian mixture models and the network
sensor localization problem, the proposed method achieves much faster convergence, with
relative error from the mean improved by about an order of magnitude than WHMC in
some cases.

\end{abstract}


\section{Introduction}
Many machine learning methods employ Bayesian analyses where the posterior probability densities are often analytically intractable. Markov Chain Monte Carlo (MCMC) is a technique admitting sampling from most posterior densities. In theory, MCMC samplers can provide unbiased sampling of posterior densities. In practice, however, Markov chains may require a rather long time to converge to the target distribution or even trapped in a local mode. 

This problem is especially acute for high dimensional multimodal target distributions
which little knowledge of the target sampling landscape is available. MCMC samples could
be drawn from just one of the modes. In high dimensional distributions that are multimodal,
it is likely that no samples are drawn from major modes. This could lead to significant bias
in the estimation of the mean and variance of the target distribution and results in false
confidence on the quality of the inference. One may think that running multiple chains
could alleviate this problem. For high dimensions, though, this may not be effective \citep{HMC}. The probability density function could be rather 
at in some subregions in a high dimensional space. This results in rather slow convergence and causes the strategy of
running multiple chains almost computationally infeasible.

%This problem is especially acute for high dimensional multimodal target distributions which little knowledge of the target sampling landscape is available. To be more specific,  MCMC samples could be drawn from just one of the modes. In high dimensional distributions that are multimodal, it is more likely that no data points were drawn from regions that have significant modes. This could lead to significant bias in the estimation of the mean and variance of the target distribution and results in false confidence on the quality of the inference. One may think that running multiple chains could alleviate this problem. For high dimensions, though, this may not be effective, see \citep{HMC}.The probability density function could be rather flat in some subregions in a high dimensional space. This results in rather slow convergence and causes the strategy of running multiple chains almost computationally infeasible.

Recent efforts have been made to sample from  multimodal distributions, such as Regenerative Darting Monte Carlo (RDMC, \citet{RDMC}) and Wormhole Hamiltonian Monte Carlo (WHMC, \citet{WHMC}). Both methods update the sampler to include new modes as they are found during regeneration times, when the Markov chains probabilistically restarts. Markov chains probabilistically jump to other modes when they enter predefined regions determined by  known modes. During regeneration times, a local search method, such as BFGS, is used to search for new modes and redefine the jump regions. It has been shown in the literature that WHMC converges rather fast if detailed and accurate knowledge about modes are provided. This requirement, however could be unrealistic in high dimensions such as 100 dimensions or higher due to the cost in discovering new modes.

%TALK ABOUT RESIDUAL POTENTIAL IN WHMC AND WHY IT IS BAD. 1) FAKE MODES. 2) DOESN"T ELIMINATE MODES EFFICIENTLY. 3) EXTRA BIAS
In order to avoid rediscovering known modes in WHMC, \citet{WHMC} employed the so called residual potential in order to diminish the basins of attraction of known modes so that BFGS is more likely to find unknown modes when the residual potential is optimized. Even though the method has shown to be successful in specific cases, such as Gaussian mixture models, there are a few issues to this approach. First, a major drawback is that there is a possibility of fictitious modes as their method directly replaces the target distribution with the residual potential to be optimized. In fact, during our experiments we discovered that fictitious modes are indeed introduced, unstabilizing WHMC. Second, the starting points are still chosen randomly in the search space for the BFGS runs. This allows BFGS to rediscover already known modes and wasting computational resources. Another major drawback is that the independent sampler employed within WHMC requires the proposal distribution to be similar to the target distribution. Otherwise, the regeneration probability would be very small and the mode search practically never starts. This creates an awkward scenerio where one requires some knowledge of the target distribution for regeneration to be effective when no such knowledge is available apriori. Lastly, WHMC samples from the target distribution ``on-the-fly'', where mode searching is conducted during sampling. This introduces an additional source of bias before major modes are found.

At first glance, modifying the objective function to subtract out contribution from known
modes seems trivial. However, such a subtraction scheme tends to create regions with
vanishing gradient if done perfectly, slowing the convergence of any gradient based search
algorithms. In differential geometry, a geodesic is a generalization of a straight line on
Riemmanian manifolds. The vanishing gradient problem can be avoided if one finds a
manifold on which geodesics correspond to straight lines in the 
at regions of the search
space and where the geodesics would pass through the maxima of the objective function.
Recently, \citet{SGEO} showed that geodesics on manifolds conformally related to
Euclidean spaces possesses such properties and a single geodesic can travel through multiple
maxima. Such geodesics appear to be an excellent candidate for finding major modes.

We propose a novel mode searching method that outperforms the residual potential method by addressing the drawbacks experienced by WHMC. Our method first constructs a conformal geodesic that is able to visit the basins of many local maxima\footnote{The same trajectory was used as a component of a global optimization algorithm in \cite{SGEO}. The novel contribution is our mode finding method as opposed to that with a residual potential.}. The trajectory travels on the {\it residual log objective function} $\phi(x) = \log(f(x)) - \log(\hat{f}(x))$, where the modes have been subtracted out from the log objective function $f(x)$. This method addresses the first two concerns mentioned above with WHMC. First,
a starting point is chosen with a bias towards undiscovered modes so that undiscovered
modes are more likely to be found. Second, because the original function is maximized,
there will be no fictitious maxima introduced. Our experiments show that our approach
requires fewer BFGS calls to for mode searching and was stable in the experiment where
WHMC was not able to converge. Lastly, we show theoretically that WHMC can be made
more effcient by having fewer samples during the early runs where not many major modes
are found. Therefore we simply force our sampler to search for new modes after a number
of samples have been acquired so as to discover new modes as fast as possible in order to
reduce the additional source of bias introduced by on-the-
y sampling.

This article is organized as follows. Section 2 provides a brief review of MCMC literature.
Our algorithm is proposed in Section 3. Numerical results are provided in Section 4. Section
5 concludes the paper with discussion and plan for future work.

\section{Related Work}
%DONE
The first MCMC sampler developed is the Metropolis-Hastings algorithm (MH) \citep{Metropolis, Hastings}. The Gibbs sampler was proposed more recently \citep{Gibbs}. It relies on a user-specified proposal function, one that is easy to sample from, to generate candidate points. Whether the candidate points are accepted is based on the acceptance probability. The Metroplis-Hastings algorithm is rather inefficient in today's standards, however. First, its computational time scales with dimensionality as $d^{2}$ \citep{Creutz} which makes it rather inefficient in high dimensions. As a consequence,  the optimal acceptance rate is just 0.23 for random walk proposal functions\footnote{For a derivaton of the optimal acceptance rate, see Section 4 of \citep{HMC} }. It also mixes poorly in colinear regions of the target distribution.

%DONE
A better alternative is the so called Hamiltonian Monte Carlo (HMC) \citep{HMC}. The candidate points are proposed by solving the Hamiltonian equations of motion under the potential energy $U(\mathbf{x)} = - \log \pi(\mathbf{x})$, where $\pi(\mathbf{x})$ is the target distribution and the kinetic energy  $ K(\mathbf{x}) = \frac{1}{2} \mathbf{p}^T M^{-1} \mathbf{p}$, where $\mathbf{p}$ is the momentum and $M$ is a constant real and symmetric mass matrix, usually set to be the identity. First, the momentum $\mathbf{p} \sim  N(0,1)$ is sampled. Then the Hamiltonian equations of motion is solved using the leapfrog integrator and a candidate point is chosen given the trajectory length and the leapfrog step size. HMC is much more efficient than MH with optimal acceptance rate at 0.65. Also, its complexity scales with dimensions as $d^{5/4}$ which is less costly than MH in high dimensions. HMC has a few drawbacks. The leapfrog step size must be randomized to avoid periodic behaviors leading to poor mixing. Furthermore, the trajectory length and leapfrog step sizes need to be tuned from a prelimineary HMC run to obtain an optimal acceptance rate.

%Recent algorithms based on HMC
In recent years, MCMC samplers outperforming HMC has been proposed. One example is the Riemannian Manifold Hamiltonian Monte Carlo (RMHMC) \citep{RHMC}. In this algorithm the trajectories follow the Hamiltonian dynamics on a manifold defined by the Fisher-Rao metric tensor \citep{RaoMetric}. In general, the Hamiltonian contains terms depending on $\mathbf{x}$ and $\mathbf{p}$ in an inseparable way. As a result the equations of motion are implicit and the solution requires a numerical intergrator such as fixed point iteration, which may be numerically unstable and computationally costly in some cases. To alleviate this, Riemannian MCMC with Lagrangian dynamics (LMC) \citep{LMC} was proposed. LMC uses a semi-explicit integrator, in which one of the updates is explicit.

%Adaptive HMCs below
Attempts have been made to tune HMC parameters adaptively. The No-U-Turn Sampler (NUTS) \citep{NUTS} uses a recusrive algorithm to propose candidate points in a wide region of the target distribution and stops the trajectory as it makes a U-turn.  The step sizes can also be adjusted adaptively in NUTS. The Adaptive Hamiltonian and Riemann Monte Carlo (AHRMC) sampler \cite{AHRMC} uses a Bayesian optimization method to tune HMC parameters.

%Multimodal HMCs below
There exists samplers that specializes in sampling from multimodal distributions. Darting Monte Carlo (DMC) \citep{DMC1, DMC2} defines regions around locations of high posterior density and attempts to jump to another mode when entered. RDMC \citep{RDMC}, a variant of DMC, was developed so that new regions can be defined when the Markov chains reaches their regeneration times. Rather than using the Fisher Information metric, Wormhole Hamiltonian Monte Carlo (WHMC) \citep{WHMC} designs wormhole metric tensors in which the distances between modes on the corresponding manifold are reduced. Mode searches are performed with BFGS at regeneration times and the wormhole metric tensors are updated.% However, the methods mentioned above requires prior knowledge of the modes for effective sampling.

%Regeneration, mode searching, updates

\section{Motivation}
\label{motivation}
We show formally the source of bias introduced by on-the-fly sampling. Suppose we are
given a set of $N$ samples $X$, the probability density can be estimated as
\[
\hat{f}(\mathbf{x}) = \frac{1}{N}\sum_{n=1}^N K(\mathbf{x}_i- \mathbf{x}),
\]
where $\mathbf{x}_i \in X$ is a sample and $K$ is the kernel. Now, suppose that the target distribution is multimodal with $K$ modes and $k \leq K$ modes are known. Assume that $X$ is obtained on-the-fly, we can partition the samples into two sets $X = X_{k<K}\cup X_K$, where $X_{k<K}$ denotes samples before all the modes are found and $X_K$ denotes samples obtained after the discovery of all modes. Similarly, $\hat{f}_{k<K}$ and $\hat{f}_K$ denotes the density estimate obtained from $X_{k<K}$ and $X_k$, respectively and $N = N_{k<K} + N_K$ their corresponding sample sizes. Since WHMC converges to a target distribution of known modes k, the true mean and variances (and higher statistical moments) are obtained from $X_{K}$ as $N_K \rightarrow \infty$.

Without the loss of generality, consider an estimation of the mean using $X$, 
\[
\hat{\mu} = \frac{1}{N}\sum_{\mathbf{x} \in X} \mathbf{x} \hat{f}(\mathbf{x}).
\] 

Partitioning $X$ as above by splitting the sum in the function estimation into two corresponding terms, we get

\[
\hat{\mu} = \frac{N_{k<K} \sum_{\mathbf{x} \in X_{k<K}}\mathbf{x}\hat{f}_{k<K}(\mathbf{x}) + N_K \sum_{\mathbf{x} \in X_K} \mathbf{x} \hat{f}_K(\mathbf{x})}{N_{k<K} + N_K}.
\]

The above expression has two terms. The first one is the bias introduced by on-the-fly sampling. In the limit $N_K \rightarrow \infty$, the first term above is negligible compared to the second and the estimated mean converges to the true mean. Another way to reduce the on-the-fly bias is to find all the major modes as soon as possible so that $f_{k<K} \simeq f_{K}$. This can be seen in the asymptotic limit ($N \rightarrow \infty$) where the sum can be replaced by an integral $\sum_{\mathbf{x}} \rightarrow \int d\mathbf{x}$. Note that in the asymptotic limit $\mathbf{x}$ is in both $X_{K}$ and $X_{k<K}$ so the distinction between each set can be dropped. Finally, set $f_{k<K} = f_{K}$ and the denominator is canceled by the corresponding term in the numerator. One obtains and estimate of the true mean

\[
\hat{\mu} = \int \mathbf{x}\hat{f}_K(\mathbf{x}) d\mathbf{x}.
\]

The above steps only involves manipulation of the density estimate and therefore is valid for all higher statistical moments. From the above consideration one sees that the on-the-fly bias can be reduced by having an efficient mode finding algorithm such that major modes first so that $\hat{f}_{k<K} \simeq \hat{f}_K$, and that limiting the number of samples before all modes are found can speed up convergence. Even though it may be impractical to find all modes before sampling, we found that the proposed algorithm, when applied on-the-fly, still outperforms WHMC.

\section{Experiments}
We perform comparisons between the tempered residual potential energy (TRPE) method of WHMC and SGEO-KDE using target distributions provided by \citet{WHMC}. To
justify the discussion in Section \ref{motivation}, the density around a mode is estimated by a Gaussian
distribution with mean and covariance being the maximizer and inverse Hessian matrix at
the mode in the first experiment involving Gaussian mixtures, so that the estimated density $f_k$ for mode k is very close to a Gaussian term in the target distribution. In the second
experiment with the sensor network localization problem, KDEs were used. We found that
SGEO-KDE converges much faster than TRPE (WHMC) in both experiments. Further,
we found no ficititious modes using SGEO-KDE whereas the residual potential method
ficititious modes are present which caused the WHMC sampler to be unstable.

The following diagnostic quantities are used to assess convergence, the relative error of
mean (REM),

\[
REM(t) = \frac{\sum_d | \hat{\mu}_d(t) - \mu_d^*|}{\sum_d \mu_d^*},
\]
where $d = \{ 1, \ldots, D\}$ denotes the components, $\hat{\boldsymbol\mu}(t)$ denotes the MCMC estimate of the mean at time $t$, and $\boldsymbol\mu^*$ is the mean of the target distribution. Similarly, the relative error of covariance (RECOV) is defined as
\[
RECOV(t) =  \frac{\sum_{ij} |\hat{\sigma}_{ij}(t) - \sigma_{ij}^*|}{\sum_{ij} \sigma_{ij}^*},
\]
where indices $i$ and $j$ donotes the components of the covariance matrix $\boldsymbol\sigma^*$ and its MCMC estimate $\hat{\boldsymbol\sigma}$. 

%Finally, we also compute the R-statistic defined by \citet{Rstat} which converges to 1 from above. Figurative speaking, The R-statistic measures the approximated ratio of variances of chains and the true variance. 

Since BFGS is the computational bottleneck, especially in high dimensions. We show the average number of BFGS attempts $N_{QN}$ as a measure of computation cost.

\subsection{Mixture of Gaussians}
We test SGEO-KDE on Gaussian mixture models used by \citet{WHMC} in this section,
the estimates are obtained from 4 MCMC chains run in parallel. The HMC acceptance rate
is tuned to be near optimal, in the range of 0.6 to 0.7. The wormhole in
uence factor is set
to be F = 0.1.

In addition to the convergence diagnostics, we also compute the average number of BFGS optimization needed to locate all the modes for the SGEO optimization algorithm and the method of tempered residual potential energy employed in WHMC. This quantity can be used as a measure for computational cost as BFGS optimization is the bottleneck for both algorithms.

\begin{figure}[t]
\center
    \includegraphics[width=1\linewidth]{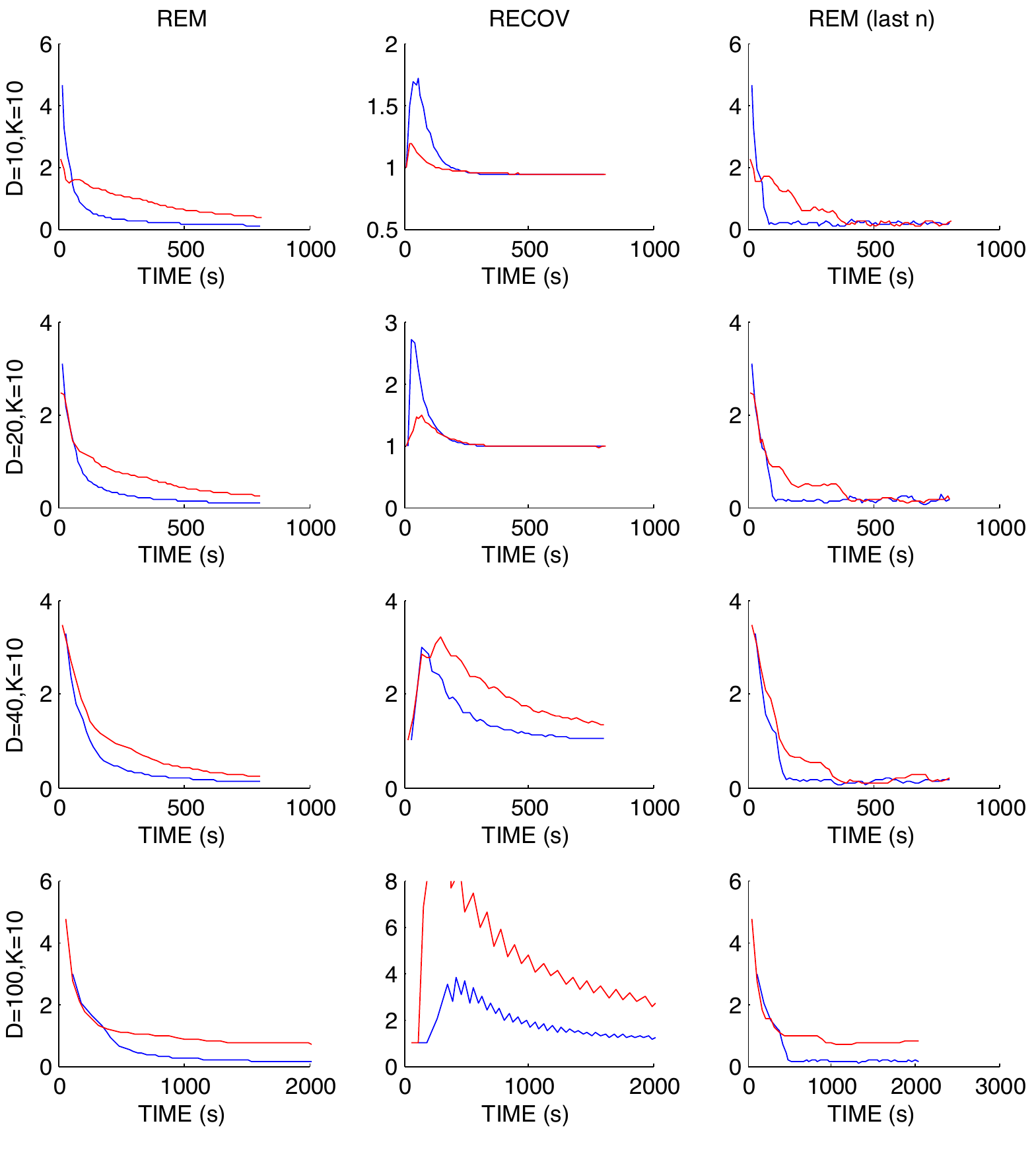}
\caption[]{Convergence diagnostics of SGEO-KDE (blue) and TRPE (WHMC) (red) for
Gaussian mixture models of equal weights. The columns represent the results for
REM, RECOV and the REM using the most recent 1000 samples, respectively.
The rows denotes the results for various mixture models.}
\label{fig:MoG1}
\end{figure}

Results for SGEO-KDE and TRPE (WHMC) for Gaussian mixture models with equal
weights are shown in Table \ref{tab:MoG1}. The corresponding results for Guassian mixtures with unequal
weights
\[
w_k \propto k/K, 
\]
where $k = \{1, \ldots, K \}$, are shown in Table \ref{tab:MoG2}. The results are taken after running the
samplers for 800 seconds, and 2000 seconds for the $D$ = 100 case. Plots of convergence
diagnostics over computational time are shown in Figure \ref{fig:MoG1} and Figure \ref{tab:MoG2} for the unequal
weight and equal weight cases, respectively. From the results we see that SGEO-KDE
outperforms TRPE (WHMC) in all diagnostic measures, with the exception of three cases
in RECOV. In addition, the SGEO mode searching algorithm is shown to be superb in
finding new modes, with only one failure out of a total 80 attempts in Tables \ref{tab:MoG1} and \ref{tab:MoG2}.
Furthermore, TRPE (WHMC) requires at least four times the number of BFGS attempts
than SGEO-KDE. This shows that SGEO-KDE outperforms TRPE (WHMC) in terms of
computational cost as the bottleneck is BFGS. This is confirmed by conisdering the REM
measure from the most recent 1000 samples (third column of Figures 1 and 2). SGEO-KDE
takes a much shorter time to converge to the true mean. The improvement is amplified as
the number of dimensions increase. In the $D$ = 100 case, TRPE (WHMC) was not able
to locate all the modes in 2000 seconds whereas the proposed method where all the modes
were discovered at about 500 seconds. Furthermore, in higher dimensions ($D \geq 40$), the
margin of improvement is larger when the Gaussian weights become unequal.

The results from HMC are included in the appendix. The HMC sampler is not able to
sample from all the modes and the convergence is significantly worse than TRPE (WHMC)
and SGEO-KDE.

%The corresponding results for Guassian mixtures with equal weights are shown in Table \ref{tab:MoG2}. The results are taken after running the samplers for 400 seconds, and 800 seconds for the $D=100$ case. Plots of convergence diagnostics over computational time are shown in Table \ref{fig:MoG1} and Table \ref{fig:MoG2} for the unequal weight and equal weight cases, respectively. Note that the time plotted takes into account of the running time of SGEO before sampling. From the results we see that SGEO-WHMC outperforms WHMC in all diagnostic measures, other than two cases in RECOV. In addition, the SGEO mode searching algorithm is shown to be superb in finding new modes, with only one failure out of a total 120 attempts in Tables \ref{tab:MoG1} and \ref{tab:MoG2}. Furthermore, the tempered residual potential method of WHMC requires at least four times the number of BFGS attempts than SGEO. This shows that SGEO-WHMC outperforms WHMC in terms of computational cost as the bottleneck is BFGS. 

%%%%%%%%%%%%%%%%%%%%%%%%%%%%%%%%%%%%%%%%%%%%%%%%%
\begin{table}[t]
\centering
\begin{tabular}{|c|c|c|c|}
\hline
$D=10, K = 10$ & REM  & RECOV  & $N_{BFGS}$\\
\hline
SGEO-KDE (all modes) & 0.018371 &0.95372   & 10\\
SGEO-KDE (on the fly) &0.083307  & 0.94318   & 10 \\
TRPE (WHMC) & 1.2962 &  3.0498 & 61.1\\
\hline
$D=20, K = 10$ &  &  &\\
\hline
SGEO-KDE (all modes) &0.030689 & 0.98234  & 10\\
SGEO-KDE (on the fly) & 0.032558  &0.98191 & 10\\
TRPE (WHMC) & 0.66759 & 3.5667 & 44.8\\
%\hline
%$D=20, K = 20$ &  &   \\
%\hline
%SGEO-WHMC (all modes) & 0.035351  & 0.99568   \\
%SGEO-WHMC (on the fly) & 0.42892  & 1.1731    \\
%WHMC &0.63643 &  0.99088 \\
\hline
$D=40, K = 10$ &  &  &\\
\hline
SGEO-KDE (all modes) & 0.021922  & 0.99505 &  10 \\
SGEO-KDE (on the fly) &0.17626&1.8093   & 10\\
TRPE (WHMC) & 0.47974 &  1.0246 & 47.9\\
\hline
$D=100, K = 10$ &  & &\\
\hline
SGEO-KDE (all modes) & 0.024986  & 0.99685 & 10  \\
SGEO-KDE (on the fly) & 0.12135 & 2.9112  & 10\\
TRPE (WHMC) & 0.82776 & 1.3427 & 62.4\\
\hline
\end{tabular}
\caption{Comparisons between SGEO-KDE and WHMC using Gaussian mixture models
with equal weights.  }
\label{tab:MoG1}
\end{table}

%%%%%%%%%%%%%%%%%%%%%%%%%%%%%%%%%%%%%%%%%%%%%%%%%
%%%%%%%%%%%%%%%%%%%%%%%%%%%%%%%%%%%%%%%%%%%%%%%%%

%%%%%%%%%%%%%%%%%%%%%%%%%%%%%%%%%%%%%%%%%%%%%%%%%
\begin{table}[t]
\centering
\begin{tabular}{|c|c|c|c|}
\hline
$D=10, K = 10$ & REM &  RECOV & $N_{BFGS} $ \\
\hline
SGEO-KDE (all modes) &0.17218&0.95234  & 10.05 \\
SGEO-KDE (on the fly) & 0.14069&0.94555   & 10.05 \\
TRPE (WHMC) & 0.9356&0.96606 & 37.5 \\
\hline
$D=20, K = 10$  & &  &\\
\hline
SGEO-KDE (all modes) & 0.14588&0.98155 &10  \\
SGEO-KDE (on the fly) & 0.10588&0.9638   &10 \\
TRPE (WHMC) & 0.68653&0.96862  & 44.2\\
%\hline
%$D=20, K = 20$ &  &   \\
%\hline
%SGEO-WHMC (all modes) & 0.10447&0.99554  \\
%SGEO-WHMC (on the fly) & 0.19072 & 1.0108    \\
%WHMC & 0.75438&0.9751 &\\
\hline
$D=40, K = 10$ &  &  &\\
\hline
SGEO-KDE (all modes) & 0.11964&0.99498 & 10   \\
SGEO-KDE (on the fly) & 0.1011 &  1.2841   & 10\\
TRPE (WHMC) &1.3478&15.659 & 43.5\\
\hline
$D=100, K = 10$  & & &\\
\hline
SGEO-KDE (all modes) & 0.07165&0.99667  &10  \\
SGEO-KDE (on the fly) & 0.078958 &  2.5587  &10 \\
TRPE (WHMC) & 0.73814&1.6344 & 61.1\\
\hline
\end{tabular}
\caption{Same as Table \ref{tab:MoG1} but using Gaussian mixture models with unequal weights $p_k \propto k/K$, where $k = \{1, \ldots, K \}$.}
\label{tab:MoG2}
\end{table}

%%%%%%%%%%%%%%%%%%%%%%%%%%%%%%%%%%%%%%%%%%%%%%%%%
\begin{figure}[t]
\center
    \includegraphics[width=1\linewidth]{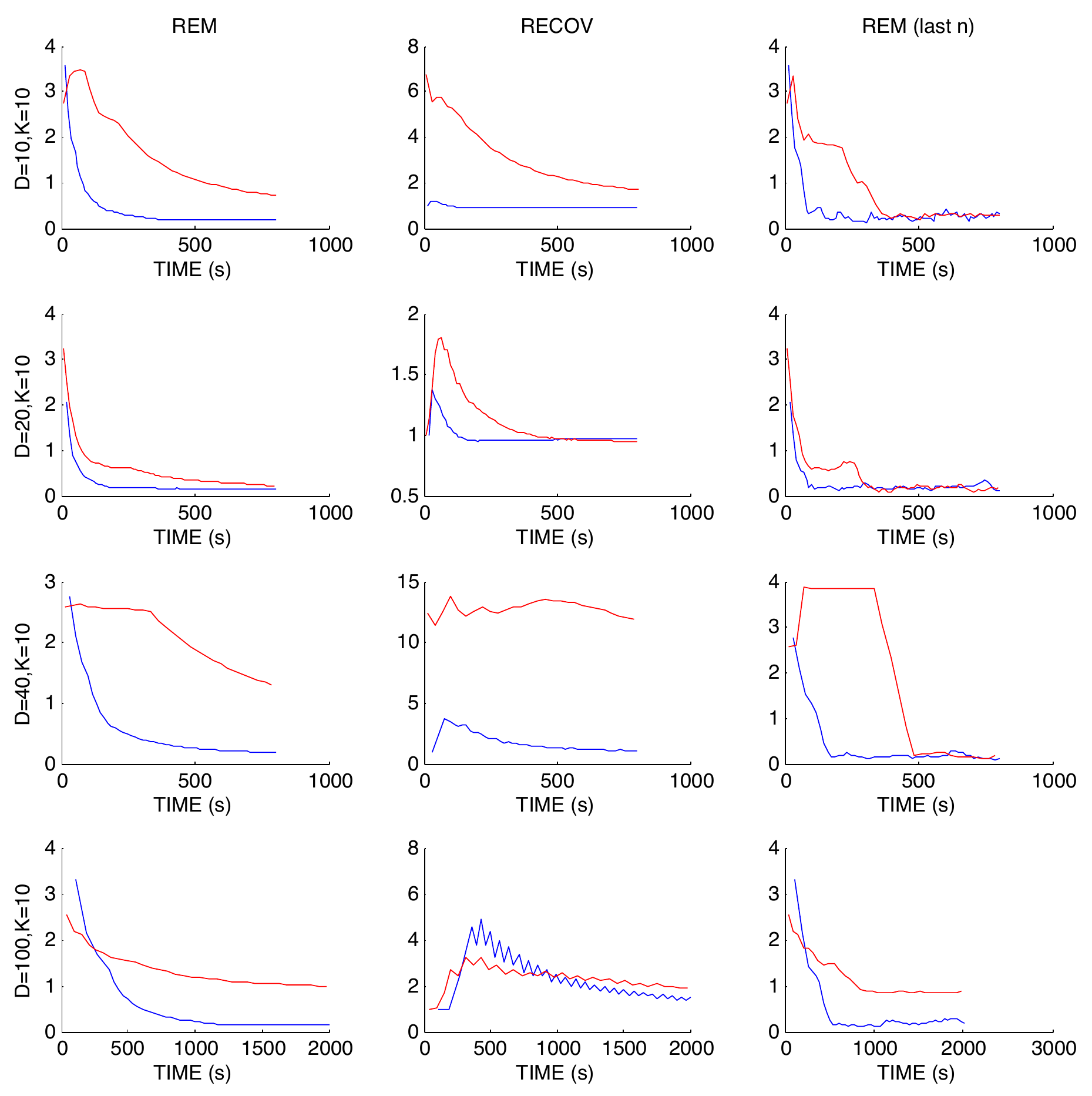}
\caption[]{Same as Figure \ref{fig:MoG1} but with unequal weights. SGEO-KDE (blue) and TRPE (WHMC) (red).}
\label{fig:MoG2}
\end{figure}
%%%%%%%%%%%%%%%%%%%%%%%%%%%%%%%%%%%%%%%%%%%%%%%%%

\subsection{Sensor Network Localization}
Given a set of $N_s$ sensors scattered on a two dimensional plane, it is possible to infer the locations of each sensor given the distance measured between each sensors. The domain of the posterior distribution is a $2N_s$ dimensional configuration space comprises of the locations of each sensor $\{ \mathbf{x}_1, \ldots, \mathbf{x}_{N_s} \}$, where each $\mathbf{x}_i$ is a 2-dimensional coordinate vector denoting the position of sensor $i$. The solution for the locations can be found by a maximum a-
posterior (MAP) configuration estimate, and its uncentainties estimated by MCMC samples
of the posterior. Due to measurement error, the posterior distribution can be multimodal,
which each mode denoting a possible set of locations for the Ns sensors that maximizes the
posterior locally. Obviously, the modes are not known aprori and hence this problem is an
excellent application for SGEO-KDE. We run two chains in parallel for this experiment.

In general, each mode of the posterior is non-Gaussian. After finding a new mode, HMC samples are taken and used to obtain a Gaussian kernel estimate of the posterior around the new mode.Then the KDE estimate of each mode is subtracted out for SGEO. Obtaining
the true mean and covariance matrix of the target distribution is intractible. Thus we
perform a long SGEO-KDE run to obtain an estimate of the ``true'' mean and covariance
of the target distribution.

If the prior distribution is taken to be uniform, the target distribution has the form given by \citep{WSN2}

\begin{eqnarray}
 \pi (\mathbf{x}_1, \ldots, \mathbf{x}_N) &=& \prod_{i>j} \psi_{ij}(\mathbf{x}_i, \mathbf{x}_j), \nonumber \\
\psi_{ij}(\mathbf{x}_i, \mathbf{x}_j) & = & [1-P_o(\mathbf{x}_i, \mathbf{x}_j)]^{(1-o_{ij})} [P_o(\mathbf{x}_i, \mathbf{x}_j)P_{\nu}(d_{ij}- ||\mathbf{x}_i - \mathbf{x}_j||)]^{o_{ij}}, \nonumber \\
P_o(\mathbf{x}_i, \mathbf{x}_j) & = & \exp(-\frac{||\mathbf{x}_i - \mathbf{x}_j||^2}{2R^2}), \qquad\textrm{and} \nonumber \\
P_{\nu}(\cdot) & = & N(\cdot |0, \sigma^2), \nonumber \\
\end{eqnarray}
where $o_{ij} = \{0,1\}$ denotes whether sensors $i$ and $j$ detect each other. If so, $d_{ij}$ denotes the distance measured between the sensors. The total number of unknown sensors is $N_s$. In accordance to the literature, we choose $N_s =8, R  = 0.3$ and $\sigma = 0.02$. The wormhole influence factor is $F=0.01$, with the probability of jumping between modes being $0.68$. The HMC parameters are set to be $\epsilon = 0.014$, $L = 0.14$ and the mass matrix $M = \mathbf{I}$. The HMC acceptance rate is 0.60. 

\begin{figure}[t]
\center
    \includegraphics[width=1\linewidth]{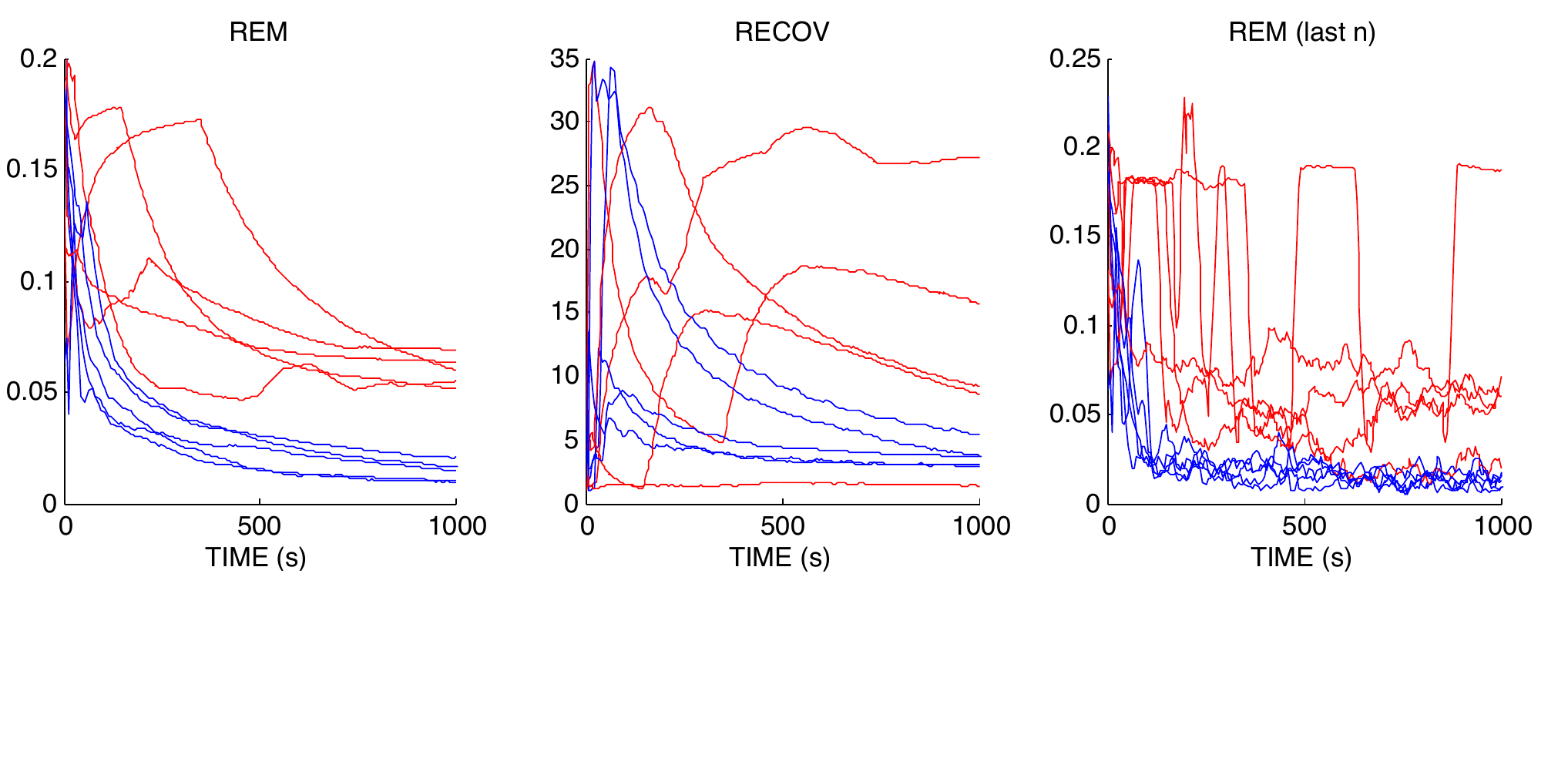}
\caption[]{Convergence diagnostics for the sensor network localization problem. SGEO-WHMC (on the fly, blue, all modes, magenta) and WHMC with forced mode updates (green).}
\label{fig:WSN}
\end{figure}

The results are shown in Table \ref{tab:WSN} and Figure \ref{fig:WSN}. In Figure \ref{fig:WSN}, it is shown that WHMC is not able to converge even after 1000 seconds in the plot on the right. For SGEO-KDE, convergence is achieved very quickly. The reason is that the tempered residual potential method finds fictitious modes that destabilize WHMC. To show this, we investigated whether another BFGS run starting from the recorded locations of the modes would lead to another point. Let $\hat{\mathbf{x}}_k^*$ be the $k$-th recorded mode ($k = 1$ being the first mode discovered and $k = K$ denotes the last mode) from either the residual potential method or SGEO-KDE and $\mathbf{x}_k^*$ be a local maximum of the target distribtion obtained by BFGS using $\hat{\mathbf{x}}_k^*$ as a starting point. Then we consider the displacement between $\hat{\mathbf{x}}_k^*$ and $\mathbf{x}_k^*$ up to the $k$-th mode as a measure for fictitious modes, that is
\[
\Delta x_k = | \hat{\mathbf{x}}_k^* - \mathbf{x}_k^*| .
\]

\begin{figure}[t]
\center
    \includegraphics[width=0.9\linewidth]{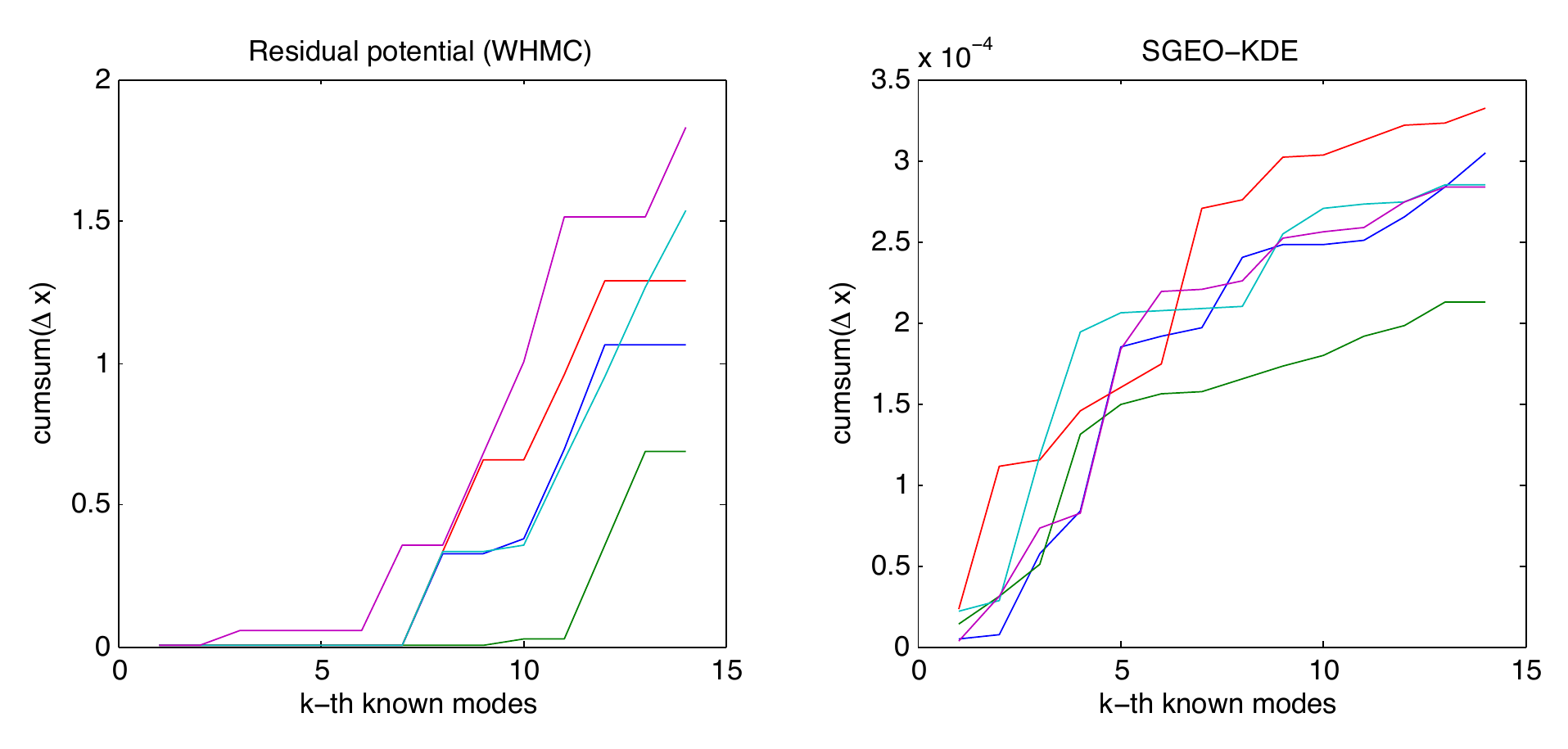}
\caption[]{This figure shows the cumulative error in mode finding, $\sum_k \Delta x_k$, as each mode
is found. The left shows the results for residual potential mode search. The
corresponding results for SGEO-KDE is shown on the right. Note the scales of
the two plots on the vertical axes.}
\label{fig:cumsum}
\end{figure}

Figure \ref{fig:cumsum} shows the cumulative displacement, $\sum_{i=1}^k \Delta x_k$ for the residual potential method and SGEO-KDE over five runs. The results show that significant error tends to develop for residual potential after some modes have been discovered. For SGEO-KDE this trend is not observed and the cumulative error remains small as new modes are being discovere. This is consistent with the fact that the residual potential introduces ficititious modes as genuine ones are being found.

%%%%%%%%%%%%%%%%%%%%%%%%%%%%%%%%%%%%%%%%%%%%%%%%%
%\begin{figure}[t]
%\center
%    \includegraphics[width=0.9\linewidth]{plots/WSN_SGEOWHMC.png}
%\caption[]{asdf}
%\label{fig:WSN_SGEOWHMC}
%\end{figure}
%%%%%%%%%%%%%%%%%%%%%%%%%%%%%%%%%%%%%%%%%%%%%%%%%

\begin{table}[t]
\centering
\begin{tabular}{|c|c|c|}
\hline
 & REM $(10^{-3})$ &  $RECOV$\\
\hline
SGEO-KDE (all modes)&7.01&0.14\\
SGEO-KDE (on the fly)&17.8&4.45\\
TRPE (WHMC)&66.4&13.93\\
\hline
\end{tabular}
\caption{Convergence diagnostics for the sensor network localization problem averaged over 5 runs corresponding to Figure \ref{fig:WSN}. Only the second half of the chains were used.}
\label{tab:WSN}
\end{table}

%%%%%%%%%%%%%%%%%%%%%%%%%%%%%%%%%%%%%%%%%%%%%%%%%
%%%%%%%%%%%%%%%%%%%%%%%%%%%%%%%%%%%%%%%%%%%%%%%%%
\section{Discussion and Conclusions}
In principle, the mode searching algorithm described in this article can be applied to many
MCMC methods without modification. We have also tried our algorithms on a Gaussian
mixture model with 500 dimensions and obtained analogous results for mode searching.
The next challenge is to extend our algorithm to much higher dimensions than described in
this article.
% Manual newpage inserted to improve layout of sample file - not
% needed in general before appendices/bibliography.

\newpage

\appendix
\section{ Review of MCMC algorithms}

\subsection{HMC}
The HMC originates from Hamiltonian dynamics. The solution of the Hamiltonian equations of motion describes the trajectories of particles under a potential energy $U(\mathbf{x})$ and the kinetic energy $K(\mathbf{p})$, where $\mathbf{x}$ is the position and $\mathbf{p}$ is the momentum of the particle. In HMC, the potential is chosen to be the negated log-density, i.e. $U(\mathbf{x}) = - \log \pi(\mathbf{x})$, where $\pi$ is the target density. The kinetic energy is $K(\mathbf{p}) = \frac{1}{2} \mathbf{p}^T M^{-1} \mathbf{p}$, where $M$ is a constant mass matrix, usually set to be the identity. Given a trajectory length $L$, each candidate point is chosen as the last position of its trajectory.

Let $H(\mathbf{x},\mathbf{p}) = U(\mathbf{x} ) + K(\mathbf{p})$ be the Hamiltonian. The Hamiltonian equations of motion are

\begin{eqnarray}
\frac{d \mathbf{x}}{dt} & = & \nabla_{\mathbf{p}} H,  \nonumber \\
\frac{d \mathbf{p}}{dt} & = & -\nabla_{\mathbf{x}} H. \nonumber 
\end{eqnarray}

The above can be solved numerically by using the leapfrog integrator with step size parameter $\epsilon$

\begin{eqnarray}
\mathbf{p}(t + \frac{\epsilon}{2}) & = & \mathbf{p}(t) - \frac{\epsilon}{2} \nabla_{\mathbf{x}} U(\mathbf{x}(t)) \nonumber \\
\mathbf{x}(t + \epsilon) & = & \mathbf{x}(t) + \epsilon M^{-1} \mathbf{p}(t + \frac{\epsilon}{2}) \nonumber \\
\mathbf{p}(t + \epsilon) & = & \mathbf{p}(t + \frac{\epsilon}{2}) - \frac{\epsilon}{2} \nabla_{\mathbf{x}} U(\mathbf{x}(t + \epsilon)). \nonumber
\end{eqnarray}

Let a state be $s = (\mathbf{x}, \mathbf{p})$. Analogous to MH, the acceptance probability $\alpha_{MH}(s \rightarrow s')= \min (1, w(s')/w(s))$, where $w = p(s) / g(s \rightarrow s')$, $w' = p(s') / g(s' \rightarrow s)$, $p(s)$ denotes the probability of being in state $s$ and $g(s \rightarrow s')$ denotes the transition kernel from state $s$ to $s'$. Since Hamiltonian dynamics is time reversible, we have $g(s \rightarrow s') = g(s' \rightarrow s)$. Then the acceptance probability is determined by the probability of being in states $s$ and $s'$. From statistical mechanics, the probability of finding a state $s$ is\footnote{We set the temperature parameter $T=1$ here. The distribution is also called the Boltzmann distribution used in Boltzmann machines.}

\[
p(s) = \frac{1}{Z} \exp [-H(s)],
\]
where $Z$ is a normalization constant. The acceptance probability for HMC is
\[
%\alpha_{HMC} = \min \{1, \exp[H(\mathbf{x}(t),\mathbf{p}(t)) - H(\mathbf{x}(t'),\mathbf{p}(t')] \}.
\alpha_{HMC} = \min \{1, \exp[H(s) - H(s')] \}.
\]

The algorithm for HMC is shown in \ref{alg:HMC}. The two input parameters are the leapfrog step size $\epsilon$ and the total number of leapfrog steps $N_{\epsilon}$. In practice, these parameters need to be tuned so that the acceptance probability is close to the optimal value of 0.65.

\begin{algorithm2e}
\caption{HMC}
\label{alg:HMC}
\begin{enumerate}
\item Given input $\mathbf{x}_t$, generate $\mathbf{p}_t \sim N(0,1)$
\item Obtain ($\mathbf{x}_{t+1}, \mathbf{p}_{t+1}$) with Leapfrog steps of size $\epsilon$ for $N_{\epsilon}$ steps.
\item Accept $\mathbf{x}_{t+1}$ with probability $\alpha_{HMC}$, discard $\mathbf{p}_{t+1}$.
\end{enumerate}
\end{algorithm2e}

Despite the efficiency of HMC for target distribution with a single mode,  the HMC trajectories can get trapped in only one of the modes when multiple are present. This causes biased sampling where only one of the modes are being sampled. Many methods exists to alleviate this problem. In the next subsection we review the state-of-the-art method for multimodal sampling using HMC, the Wormhole Monte Carlo (WHMC) sampler. Tables \ref{tab:HMC1} and \ref{tab:HMC2} shows the results with HMC on Gaussian mixture models.

\begin{table}[t]
\centering
\begin{tabular}{|c|c|c|}
\hline
 & REM  & RECOV  \\
\hline
$D=10, K = 10$ & 5.471  & 7.72    \\
\hline
$D=20, K = 10$ & 4.442  & 17.09   \\
\hline
$D=20, K = 20$ & 4.917  & 3.835  \\
\hline
$D=40, K = 10$ & 3.707  & 27.55 \\
\hline
$D=100, K = 10$ & 4.455  & 56.74 \\
\hline
\end{tabular}
\caption{Same as Table \ref{tab:MoG1} with results obtained from HMC.}
\label{tab:HMC1}
\end{table}

\begin{table}[t]
\centering
\begin{tabular}{|c|c|c|}
\hline
 & REM &  RECOV  \\
\hline
$D=10, K = 10$ & 5.582  & 9.921    \\
\hline
$D=20, K = 10$ & 5.068  & 12.62   \\
\hline
$D=20, K = 20$ & 4.767  & 3.028  \\
\hline
$D=40, K = 10$ & 4.742  & 18.81 \\
\hline
$D=100, K = 10$ & 4.956  & 58.39 \\
\hline
\end{tabular}
\caption{Same as Table \ref{tab:MoG2} with results obtained from HMC.}
\label{tab:HMC2}
\end{table}

\subsection{WHMC}
When multiple modes exist in the target distribution, the samples and HMC trajectories can be confined within a local maximum.  Given the locations of the modes, the WHMC employs the HMC sampler extended onto a Riemannian manifold, endowed with a metric that shortens the intermodal distances. Following \citet{RHMC}, the constant mass matrix $\mathbf{M}$ in the kinetic energy is replaced by the positional dependent metric, $\mathbf{M} = \mathbf{G}(\mathbf{x})$. An immediate consequence of this is that, as we shall see, at least one of the leapfrog update equations become implicit and the updates require fixed point iterations.

Consider we have two known modes at $\mathbf{x}_1^*$ and $\mathbf{x}_2^*$. The metric in WHMC comprises of two terms, 
\[
\mathbf{G} = (1 - m(\mathbf{x})) \mathbf{I} + m(\mathbf{x}) \mathbf{G}_W,
\]
where $\mathbf{G}_W$ is the wormhole metric defined by
\[
\mathbf{G}_W = \mathbf{I} - (1-\epsilon_W) \mathbf{v}_W (\mathbf{v}_W)^T,
\]
with $\mathbf{v}_W$ being a unit vector point from $\mathbf{x}_1^*$ to $\mathbf{x}_2^*$ and $\epsilon_W$ is a small and positive number $\epsilon_W \ll 1$. The mollifying function $m(\mathbf{x}) \in (0,1)$ is defined by
\[
m(\mathbf{x}) = \exp\{-\frac{1}{F} ( ||\mathbf{x} - \mathbf{x}_1^*|| + ||\mathbf{x}_2^* -\mathbf{x}|| - || \mathbf{x}_2^* - \mathbf{x}_1^*||) \},
\]
where $F$ is the wormhole influence factor.

Notice that, using the triangle inequality,
\[
||\mathbf{x} - \mathbf{x}_1^*|| + ||\mathbf{x}_2^* -\mathbf{x}|| \geq || \mathbf{x}_2^* - \mathbf{x}_1^*||,
\]
$m(\mathbf{x}) = 1$ if and only if $\mathbf{x}$ is on the line connecting the two modes and that the wormhole influence factor controls the extent the wormhole metric $\mathbf{G}_W$ influences the metric $\mathbf{G}$ away from the intermodal line. Also, the wormhole metric $\mathbf{G}_W$ shortens the distance between the two modes. Let $\mathbf{v} = k \mathbf{v}_W$, where $k =|| \mathbf{x}_1^* - \mathbf{x}_2^*|| $ so that $||\mathbf{v}|| = k$ in Euclidean space. Under $\mathbf{G}_W$, the 2-norm of $\mathbf{v}$ is
\[
\mathbf{v}^T \mathbf{G}_W \mathbf{v} = \epsilon_W k^2.
\]
That is to say that the wormhole metric shortens the distance between two modes by a factor of $\sqrt{\epsilon_W}$. For more than two modes, a network of wormholes connecting each mode is built.

To facilitate moving between modes, \citet{WHMC} introduced a vector field, $\mathbf{f}$, along the intermodel line, with norm proportional to the mollifying function and the projection of the velocity\footnote{The definition of the velocity in classical mechanics is $\mathbf{v} = \mathbf{M}^{-1}\mathbf{p}$ and we are identifying the mass matrix with the metric here.} $\mathbf{v} = \mathbf{G}(\mathbf{x})^{-1} \mathbf{p}$ onto the intermodal line
\[
\mathbf{f}(\mathbf{x},\mathbf{v}) = m(\mathbf{x}) \langle \mathbf{v}, \mathbf{v}_W \rangle \mathbf{v}_W
\]
with the corresponding modification of the Hamilton equation
\[
\frac{d \mathbf{x}}{dt} = \nabla_{\mathbf{p}} H  + \mathbf{f}(\mathbf{x},\mathbf{v}).
\]
Lastly, \citet{WHMC} also introduced an auxiliary dimension to prevent interference between the wormholes and the HMC dynamics.
 
The generalized leapfrog integrator for WHMC is
\begin{eqnarray}
\mathbf{v}(t+ \frac{1}{2})&=& \mathbf{v}(t) - \frac{\epsilon}{2} \nabla_{ \mathbf{x}} U( \mathbf{x}(t)) \nonumber\\
 \mathbf{x}(t + 1) &=&  \mathbf{x}(t) + \epsilon \bigg\{ \mathbf{v}(t+\frac{1}{2}) + \frac{1}{2} \bigg[\mathbf{f}\bigg(\mathbf{x}(t),\mathbf{v}(t+\frac{1}{2})\bigg) + \mathbf{f}\bigg(\mathbf{x}(t+1),\mathbf{v}(t+\frac{1}{2})\bigg)\bigg]\bigg\} \nonumber \\
\mathbf{v}(t +1 ) &=& \mathbf{v}(t + \frac{1}{2}) - \frac{\epsilon}{2} \nabla_{\mathbf{x}} U(\mathbf{x}(t+1)). \nonumber
\end{eqnarray}
We can see that the update equation for $\mathbf{x}(t+1)$ is implicit and fixed point iteration is needed for this update.

\subsection{Updating Target Distribution During Regeneration Time}
Regeneration times are periods where a Markov chain restarts itself probabilistically. States after regeneration is independent of states prior to the regeneration time. In thoery, another set of sampler parameters can be used for the sampler while leaving the MCMC estimates unaffected. In addition, \citet{WHMC} and \citet{RDMC} performs BFGS at regeneration times to search for new modes and updates the proposal distribution on-the-fly.

Regeneration is checked at each step. The regeneartion probability is
\[
r(\mathbf{x}_t, \mathbf{x}_{t+1}) = \frac{S(\mathbf{x}_t) Q(\mathbf{x}_{t+1})}{T(\mathbf{x}_{t+1} | \mathbf{x}_t)},
\]
where
\begin{eqnarray}
S(\mathbf{x}_t) & = &  \min \bigg\{1, \frac{c}{\pi(\mathbf{x}_t) / q(\mathbf{x}_t)}  \bigg\} \nonumber \\
Q(\mathbf{x}_{t+1}) & = & q(\mathbf{x}_{t+1})  \min \bigg\{1, \frac{\pi(\mathbf{x}_{t+1}) / q(\mathbf{x}_{t+1})}{c}  \bigg\}\nonumber \\
T(\mathbf{x}_{t+1} | \mathbf{x}_t) & = &  q(\mathbf{x}_{t+1})  \min \bigg\{1, \frac{\pi(\mathbf{x}_{t+1}) / q(\mathbf{x}_{t+1})}{\pi(\mathbf{x}_t) / q(\mathbf{x}_t)} \bigg\},\nonumber 
\end{eqnarray}
$q(\cdot)$ is the proposal indepdence distribution. Note that, $r = 1$ if and only if $q(\cdot ) = \pi(\cdot)$. Regeneration probability is dependent on the choice of the independence proposal function, and the choice of the constant $c$. If the choice of the indenpendence sampler is poor, such as in the case where only a minority of the modes in a multimodal target distribution are known, then the regeneration probability can be too small that BFGS is never triggered to find new modes. We encountered such a scenario when  testing the sensor network localization problem with WHMC.

Another problem with updating the target distribution on-the-fly is that sample estimates are biased by the samples from the previous target distribution.  In the following sections, we show that the mean and covariance estimates of the target distribution are significantly more accurate if all modes of the target distribution are found first. To this end we employ a global optimization algorithm SGEO modified to discover all the modes before sampling.

% Acknowledgements should go at the end, before appendices and references
\acks{The research of second and third authors are supported by National Science and Engineering Research Grants. We thank Shiwei Lan for providing the code for WHMC.}

\vskip 0.2in
%\bibliography{sample}

\end{document}